\documentclass[a4paper,
               keeplastbox,   % flushend option: not to un-indent last line in References
               ]{jacow}
\usepackage{pdfpages,multirow,ragged2e} %
\makeatletter%
	\ifboolexpr{bool{xetex}}
	 {\renewcommand{\Gin@extensions}{.pdf,%
	                    .png,.jpg,.bmp,.pict,.tif,.psd,.mac,.sga,.tga,.gif,%
	                    .eps,.ps,%
	                    }}{}
\makeatother

\ifboolexpr{bool{xetex} or bool{luatex}} % test for XeTeX/LuaTeX
 {}                                      % input encoding is utf8 by default
 {\usepackage[utf8]{inputenc}}           % switch to utf8

\usepackage[USenglish]{babel}

\ifboolexpr{bool{jacowbiblatex}}%
 {%
  \addbibresource{jacow-test.bib}
  \addbibresource{biblatex-examples.bib}
 }{}
\usepackage[pdfencoding=auto]{hyperref}
\hypersetup{%
 bookmarksnumbered=true,%
 colorlinks=true,%
 pdftitle={Beam Operation for Particle Physics and Photon Science with Pulse-to-Pulse Modulation at KEK injector LINAC},%
 pdfauthor={Kazuro Furukawa},%
 pdfsubject={ICALEPCS 2023 TUPDP046},%
 pdfkeywords={SuperKEKB Injector LINAC\000\012Multidisciplinary injection operation\000\012}}

\begin{document}

%1% \title{preparation OF papers for \NoCaseChange{JACoW} conferences\thanks{Work supported by ...}}
%%\title{Beam Injection Operation for Particle Physics and Photon Science Experiments \protect\\
%%with Pulse-to-Pulse Beam Modulation at KEK injector linac}
\title{Beam Operation for Particle Physics and Photon Science  % \protect\\
with Pulse-to-Pulse Modulation % \protect\\
at KEK injector LINAC}

%1% \author{A. N. Author\thanks{email address}, H. Coauthor, Name of Institute or Affiliation, City, Country \\
%1% 		P. Contributor\textsuperscript{1}, Name of Institute or Affiliation, City, Country \\
%1% 		\textsuperscript{1}also at Name of Secondary Institute or Affiliation, City, Country}
\author{
\parbox{\linewidth}{\centering
K. Furukawa\thanks{kazuro.furukawa@kek.jp}, M. Satoh, Injector LINAC group,\\
% High Energy Accelerator Research Organization (KEK), 
KEK, Tsukuba, Japan, 
% the Graduate University for Advanced Studies (Sokendai), 
SOKENDAI, Tsukuba, Japan}}

% Kazuro Furukawa, Mitsuo Akemoto, Dai Arakawa, Yoshio Arakida, Yusei Bando, Hiroyasu Ego, Yoshinori Enomoto, Toshiyasu Higo, Hiroyuki Honma, Naoko Iida, Kazuhisa Kakihara, Takuya Kamitani, Hiroaki Katagiri, Masato Kawamura, Shuji Matsumoto, Toshihiro Matsumoto, Hideki Matsushita, Katsuhiko Mikawa, Takako Miura, Fusashi Miyahara, Hiromitsu Nakajima, Takuya Natsui, Yujiro Ogawa, Satoshi Ohsawa, Yuichi Okayasu, Takao Oogoe, Muhammad Abdul Rehman, Itsuka Satake, Masanori Satoh, Yuji Seimiya, Tetsuo Shidara, Akihiro Shirakawa, Hirohiko Someya, Tsuyoshi Suwada, Madoka Tanaka, Di Wang, Yoshiharu Yano, Kazue Yokoyama, Mitsuhiro Yoshida, Takashi Yoshimoto, Rui Zhang, Xiangyu Zhou

\maketitle

\begin{abstract}
The electron and positron accelerator complex at KEK offers unique experimental opportunities in the fields of elementary particle physics with SuperKEKB collider and photon science with two light sources. In order to maximize the experimental performances at those facilities the injector LINAC employs pulse-to-pulse modulation at 50 Hz, injecting beams with diverse properties. The event-based control system effectively manages different beam configurations. This injection scheme was initially designed 15 years ago and has been in full operation since 2019. Over the years, quite a few enhancements have been implemented.  As the event-based controls are tightly coupled with microwave systems, machine protection systems and so on, their modifications require meticulous planning. However, the diverse requirements from particle physics and photon science, stemming from the distinct nature of those experiments, often necessitate patient negotiation to meet the demands of both fields.  This presentation discusses those operational aspects of the multidisciplinary facility. 
\end{abstract}

%%%

%%%%%%%%%%%%%%%%%%%%
\section{Introduction}
%%%%%%%%%%%%%%%%%%%%

Particle accelerators require large resources with advanced technologies and experienced personnel in order to construct and to operate them.  Therefore, beams from certain accelerators have been shared between several different purposes.  A long beam switching interval may be accepted between some accelerators.  On the other hand, in order to switch beams frequently the mechanism called pulse-to-pulse beam modulation (PPM) have been developed and employed for advanced accelerators~\cite{ppm-pac77,ppm-ical91}. 
However, different beam users may request variety of beam properties as well as conflicting operational concepts.  Thus, meticulous arbitration for multidisciplinary operation would be necessary.  

\begin{figure}[b]
\centering
\includegraphics[width=0.99\columnwidth]{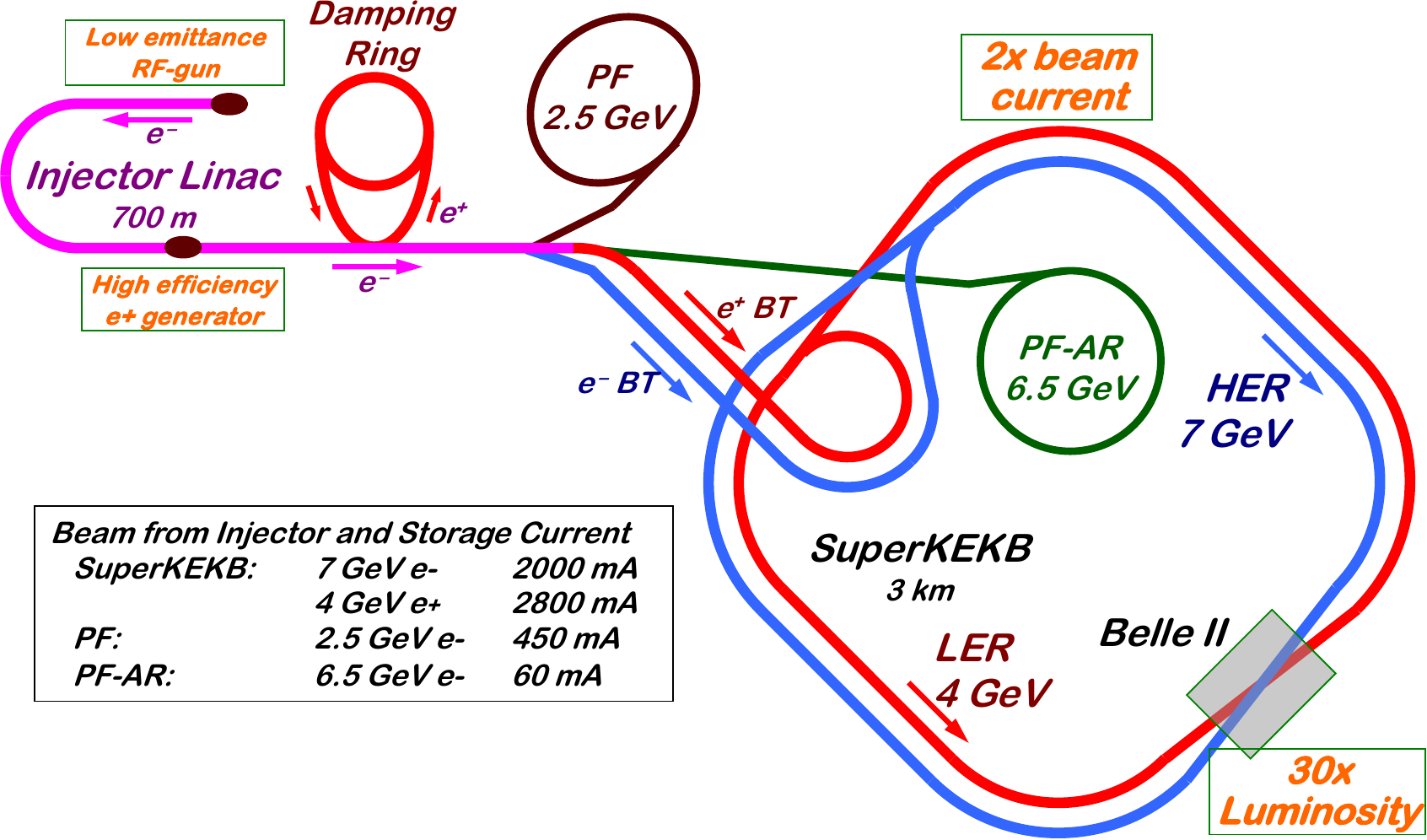}
\caption{
%%% caption Fig.1
Beam injection from electron positron injector LINAC to SuperKEKB dual ring collider as well as PF and PF-AR light source storage rings.
}
\label{skekb-diagram}
\end{figure}

The injector LINAC at KEK has been operated for more than 40 years and has served for several storage ring accelerators in the past~\cite{20khours-ipac22}.  Presently, it makes multidisciplinary injection operation to support two light sources and a B factory collider storage rings with pulse-to-pulse beam modulation as in Fig.~\ref{skekb-diagram}.  It is often referred to as simultaneous top-up injections~\cite{event-ical09}.  

%%%%%%%%%%%%%%%%%%%%
\section{Injector LINAC at KEK}
%%%%%%%%%%%%%%%%%%%%

\begin{figure*}[!th]
\centering
\includegraphics[width=0.95\textwidth]{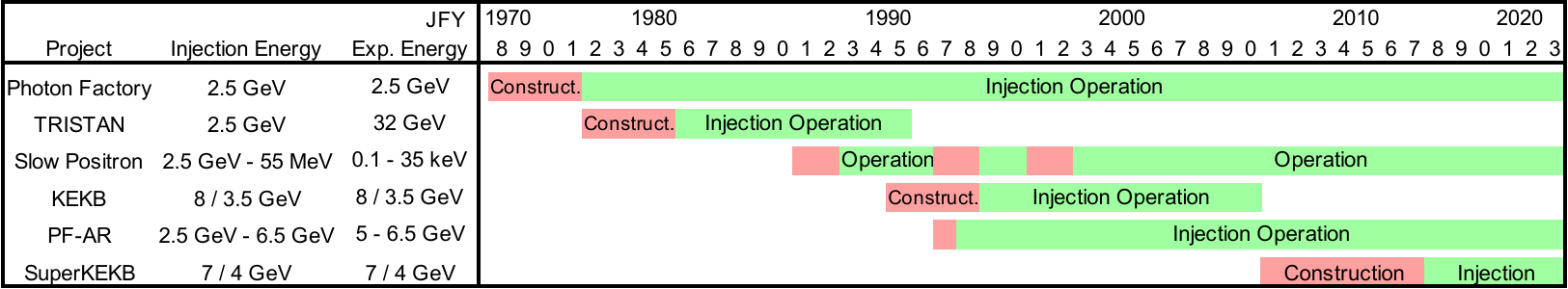}
\caption{
%% caption Fig.2
Electron positron accelerator projects operated by the injector LINAC for past 40 years.
The injector has continued the beam delivery through several upgrades since 1982 without a long shutdown. 
}
\label{linac-projects}
\end{figure*}

In 1978 a dedicated (the second generation) light source project was approved, and then a high energy electron positron collider project was approved as well.  
A 400-m electron linear accelerator was constructed to support both synchrotron radiation research and high energy physics collier experiment~\cite{linac-1980}.  Since then, it has been serving for various projects which required diverse beam properties as in Fig.~\ref{linac-projects}.  

Additionally, each project required an upgrade of the injector LINAC which might take up to six years in between projects, however, even during the upgrade the other project needed the beam injection.  Thus, the equipment tests, fabrications and installations had to be carefully arranged in order to support a sustainable injector operation. 

For example, it took 5 years to upgrade the accelerators to realize a KEKB B-factory machine~\cite{linac-kekb}.  While the collider ring was shutdown for 5 years, the injector LINAC had to deliver the beam every year to the light source.  The injector performed 3-month injection operation and 3-month upgrade installation repeatedly.  % The light source got a high brilliance upgrade approved as well, and 9-month shutdown was performed once.  
The SuperKEKB upgrade required 6 years from 2010 and another year in 2017~\cite{skb-nim2018}.  The injector utilized the last downstream part of LINAC to inject the beam for the light sources.  And the rest of the LINAC was reconstructed to support high intensity and low emittance injection for SuperKEKB as well as to recover from the Great East Japan Earthquake in 2011.  The longest complete shutdown of the injector LINAC was 5 months in 2017.  During the upgrade period a particular effort was made for the alignment in order to suppress the wakefield effect in the accelerating structure.  

In such way, the injector LINAC has negotiated with both communities of photon science and particle physics, and they understood each other to maximize the overall experimental performance.

%%%%%%%%%%%%%%%%%%%%
\section{Pulse-to-pulse beam modulation at KEK injector LINAC}
%%%%%%%%%%%%%%%%%%%%

At the beginning of KEKB project, the injector LINAC filled the storage rings once in about an hour~\cite{furukawa-linac02}.  In the later phase of KEKB it became important to perform top-up injections into both KEKB and PF experiments to maintain the stable experimental conditions.  The stable stored beam currents in the collider rings were essential to keep the collision with crab crossing by crab cavities.  At the same time the photon experiments requested top-up injection for higher precision data acquisition.  

Thus, a concept has been introduced of which a single injector behaving as multiple virtual accelerators for corresponding storage rings.  

\subsection{Simultaneous Top-up Injections}

The injector LINAC can accelerate beam bunches 50 times a second.  If device parameters for microwave generators, magnets and guns are modulated every 20 milliseconds, a single injector can perform top-up injections into multiple storage rings as if it injects beams simultaneously.  Such simultaneous injections have been realized since 2009~\cite{event-ical09}. %,simul-linac10}.  

The system is controlled globally and synchronously by an event-based PPM control system~\cite{event}.  In the system ten virtual beam modes are defined, and four of them were corresponding to injections into storage rings of SuperKEKB-HER, SuperKEKB-LER, PF and PF-AR.  Each beam mode is accompanied with about ten fast event codes, and each event code can distribute to whole accelerator complex with a single byte information and precise timing of approximately 10 picoseconds.  More than 200 device parameters along the accelerator are modulated in response to one of those event codes.  

Typical injections per second are presently 5$\sim$10 times for SuperKEKB electron, 10$\sim$25 times for SuperKEKB positron and 0.5 times for PF and PF-AR.  They are frequently adjusted by software or human operators depending on the injection status. 

\subsection{Virtual Accelerator Concept}

\begin{figure}[b]
   \centering
   \includegraphics*[width=75mm]{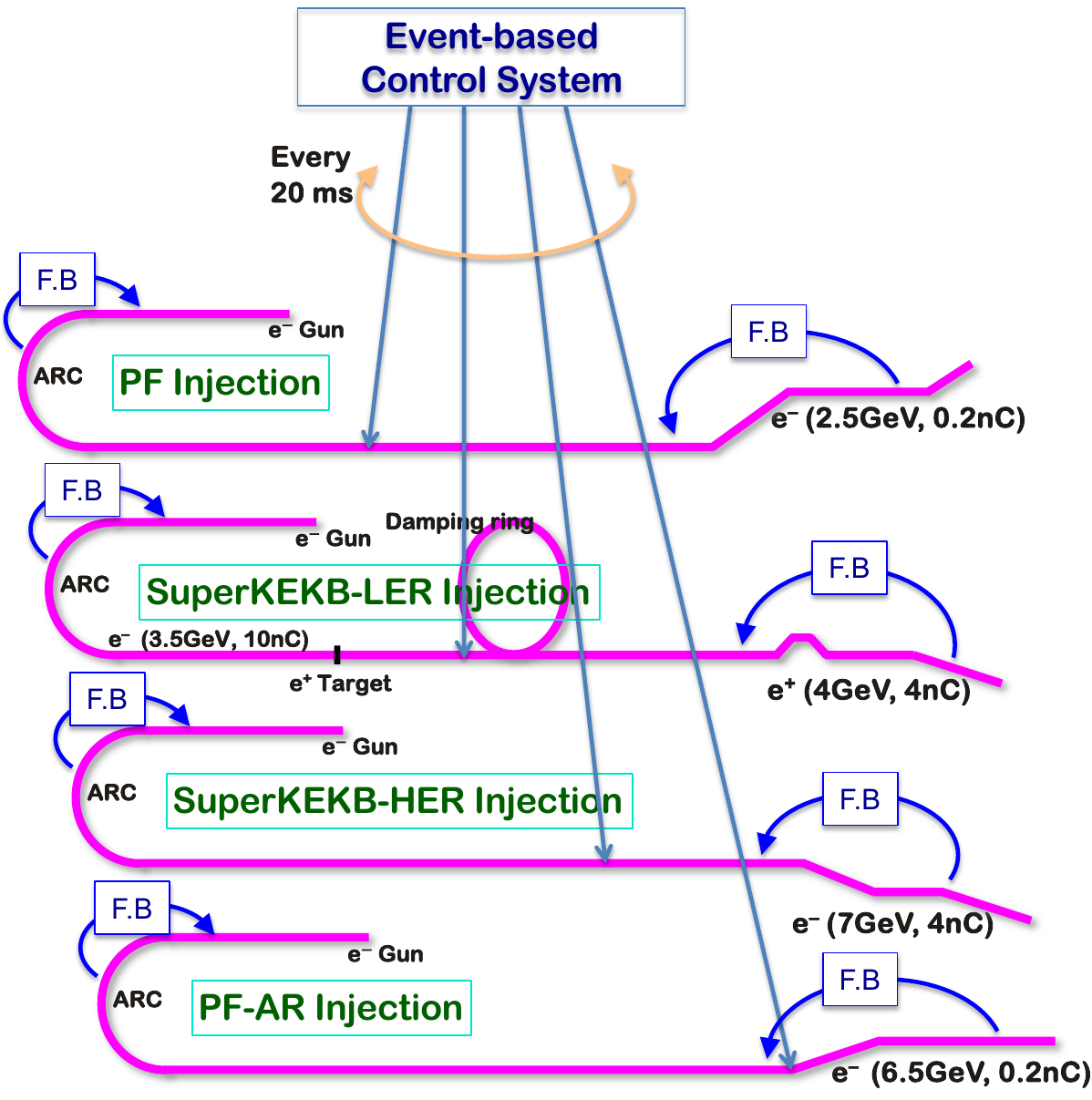}
   \caption{Single LINAC behaves as four virtual accelerators (VAs) to inject their beams into four separate storage rings. Each VA would be associated with several independent beam feedback loops.}
   \label{va}
\end{figure}

The event-based PPM controls enables not only pulse-to-pulse device controls but also pulse-to-pulse beam diagnosis readouts.  Device controls and beam diagnosis under different beam modes are completely independent.  For example, a beam stabilization feedback loop at the same physical location can behave independently under different beam modes~\cite{furukawa-ipac10}.  Such a behavior of the single injector can be recognized as multiple virtual accelerators as in Fig.~\ref{va}~\cite{control-ptep}.

As such virtual accelerators were realized, a conflict of beam requests from particle physics and photon science experiments with diverse beam qualities has been much relaxed.

%%%%%%%%%%%%%%%%%%%%
\section{Multidisciplinary Beam Injection}
%%%%%%%%%%%%%%%%%%%%

While technical issues of multi-purpose beam sharing were mostly resolved with simultaneous top-up injections, operational demanding approaches including maintenance plan and upgrade strategy must be negotiated between injector LINAC, SuperKEKB and light source divisions.  

%%%%%%%%%%%%%%%%%%%%
\subsection{Operation and Maintenance Consideration}
%%%%%%%%%%%%%%%%%%%%

The injection concepts and beam property requirements are quite different between downstream accelerators.  Because the injector LINAC is expected to support such very diverse experimental requests, the time and budget allocation optimization should be balanced between them.  

The overall schedule for the electron and positron accelerator complex is discussed between divisions three times a year.  Dates of maintenance and weekly operational details are discussed between operational leaders.  Long-term upgrade plans are developed balancing the possible outcome and risk at each accelerator.  During such discussion each division tends to focus on different aspects.   

%%%%%%%%%%%%%%%%%%%%
\subsection{Operation for Photon Science}
%%%%%%%%%%%%%%%%%%%%

3000 photon science users per year come to light source facilities in small groups for only a few days at a time.  Those short-term users require the beam operation stability over the peak performance.  If an unexpected beam shutdown occurs during the allocated experimental time slot, they may lose the opportunity to acquire experimental data. 

Thus, the risk of beam interruption must be avoided, and a four-hour routine preventive maintenance period is scheduled every other week\footnote{As the availability improved, the maintenance period will be allocated every four weeks from the autumn of 2023.}.  Light source users don't interact much with accelerator operations as far as the beam availability is satisfied.  

As the beam lifetime is more than a day, the injection beam charge per pulse is low and the injection frequency is also low as described above.  A lower value is preferable to maintain the injection radiation suppressed.  

However, from the viewpoint of operator skills, a stable operation means that operators rarely face unusual situations if they operate only on light sources. Then, they have to learn through training documents. 

%%%%%%%%%%%%%%%%%%%%
\subsection{Operation for Collider Experiment}
%%%%%%%%%%%%%%%%%%%%

On the other hand, the particle physics experiment at SuperKEKB is essentially performed by a large single user group, Belle II.  It always pursues the limit of performance.  They expect a highest possible volume of experimental collision data within a year for higher statistics.  

They often take risks of a short-term beam interruption if the integrated performance might increase.  They are reluctant to allocate time for preventive maintenance period.  They hope to continue their experiment until an accelerator device or the beam fails.  They are sometimes involved in machine improvement and machine-detector development.  They aren't afraid to take a chance to change the machine and to improve the performances.  

The stored beam current has been gradually increased more than an ampere, and the beam lifetime is typically less than 10 minutes.  Even though the beam charge per injection bunch is more than 2 nC, beam injection frequencies for electron and positron are more than 10 Hz respectively as described above.  As the detector data acquisition is presently forced to suspend for about a millisecond during the injection, they prefer larger beam charge.   There exist many challenging issues related to such large beam charges, that must be solved one by one. 

When the operators deal with the beam for particle physics, they often face small failures because of such improvement challenges. Thus, they continue to receive on-the-job training.  

%%%%%%%%%%%%%%%%%%%%
\subsection{Continuous Improvements and Protection Systems}
%%%%%%%%%%%%%%%%%%%%

During the discussion between accelerator divisions, additional beam control requirements may arise, that would be later implemented.  As the injection system has become complicated, such modification needs careful considerations.  

The safety systems especially require thorough investigations to make them work without any failures~\cite{safety-ipac13}.  For example, the interlock threshold of beam charge for radiation limit is hundred times different depending on the beam injection modes or the beam dumps~\cite{chargelimit-rsi}.  The beam repetition is 50 times a second and the beam charge need to be integrated for each beam mode.  The measurement and beam shutdown have to be performed precisely.

%%%%%%%%%%%%%%%%%%%%
\section{Operational Achievements}
%%%%%%%%%%%%%%%%%%%%

Before the construction of the injector, there was growing expectation for the world-class Japanese domestic collider in the experimental particle physics research after the successes in the world-level theoretical research.  On the other hand, in the field of synchrotron radiation science, there were demands for a dedicated accelerator for synchrotron radiation research. 

In order to realize both of those accelerator projects, an electron linear accelerator was constructed in 1982. % as part of the previously approved Photon Factory (PF), a synchrotron radiation experimental facility.
Since then, the injector LINAC has been in operation for more than 40 years, supporting the consecutive accelerator projects of Photon Factory (PF), TRISTAN, KEKB, PF-AR, and SuperKEKB as depicted in Fig.~\ref{linac-projects}.  The injector facility also runs the slow positron facility for long that also contributes to the material science field~\cite{slowpos-wada2018}. 

\begin{figure}[!htb]
   \centering
   \includegraphics*[width=\columnwidth]{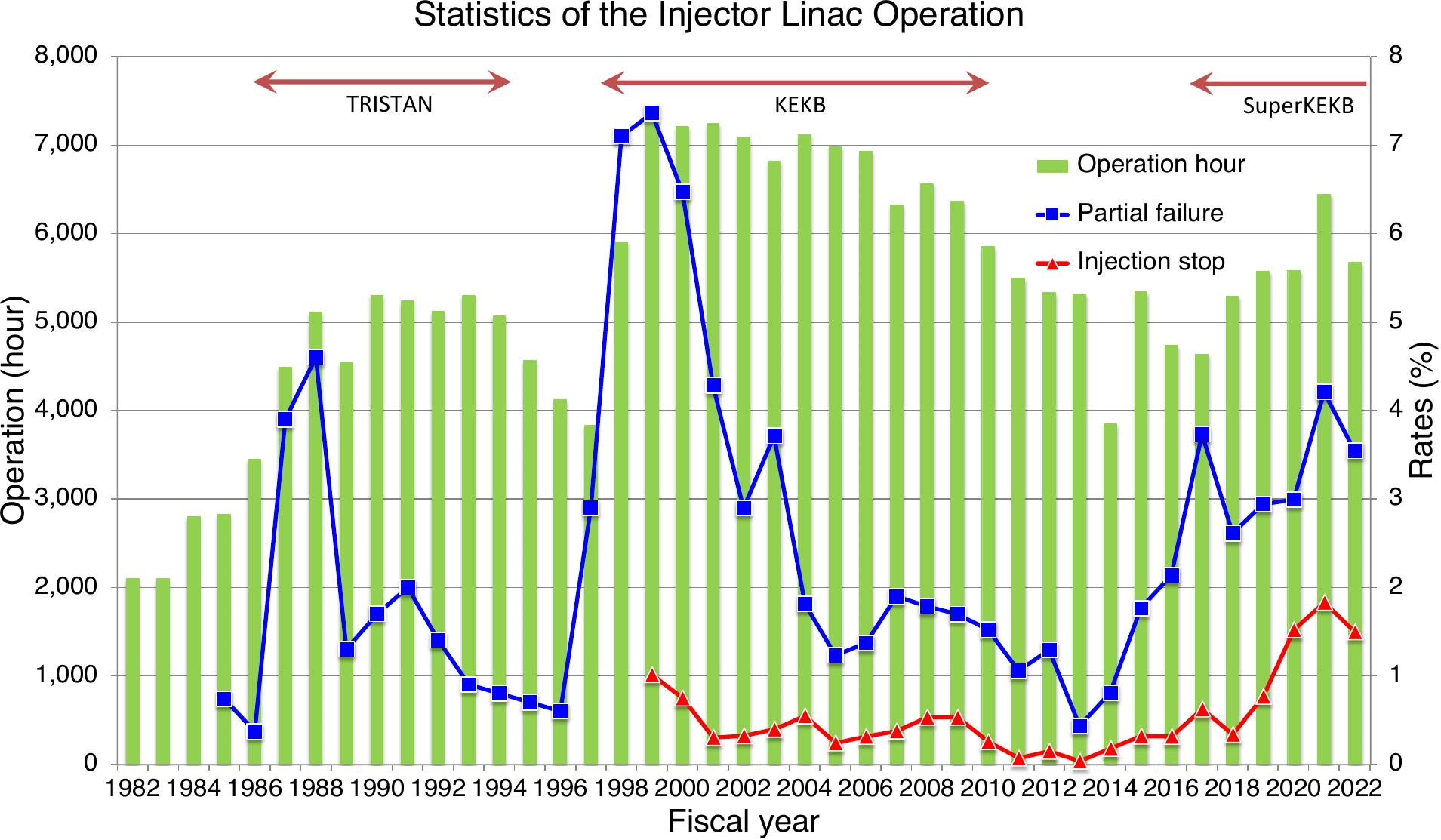}
   \caption{Yearly operation hours and failure rates of the injector LINAC. The blue line indicates device failure rate, that could have been covered by redundancy and so on.  The red line indicates the complete beam interruption. 
   }
   \label{linac-stat}
\end{figure}

The brief operation history is given in Fig.~\ref{linac-stat}, that shows long-term yearly operation hours without a long break.  The beam availability has been mostly kept more than 99\% depending on the control and operational changes in the year.  

%Continuous improvements were applied to the injector 

The continuous operation of the electron-positron injector LINAC has supported the history of advanced research at KEK by providing beams for various projects as shown in Fig.~\ref{linac-projects}.

Recent introduction of event-based PPM controls that enabled simultaneous top-up injections for multiple purposes was most significant since the construction of the injector.  That demonstrates how the controls contribute to the performance of accelerators.  

\begin{figure}[h]
\centering
\includegraphics[width=\columnwidth]{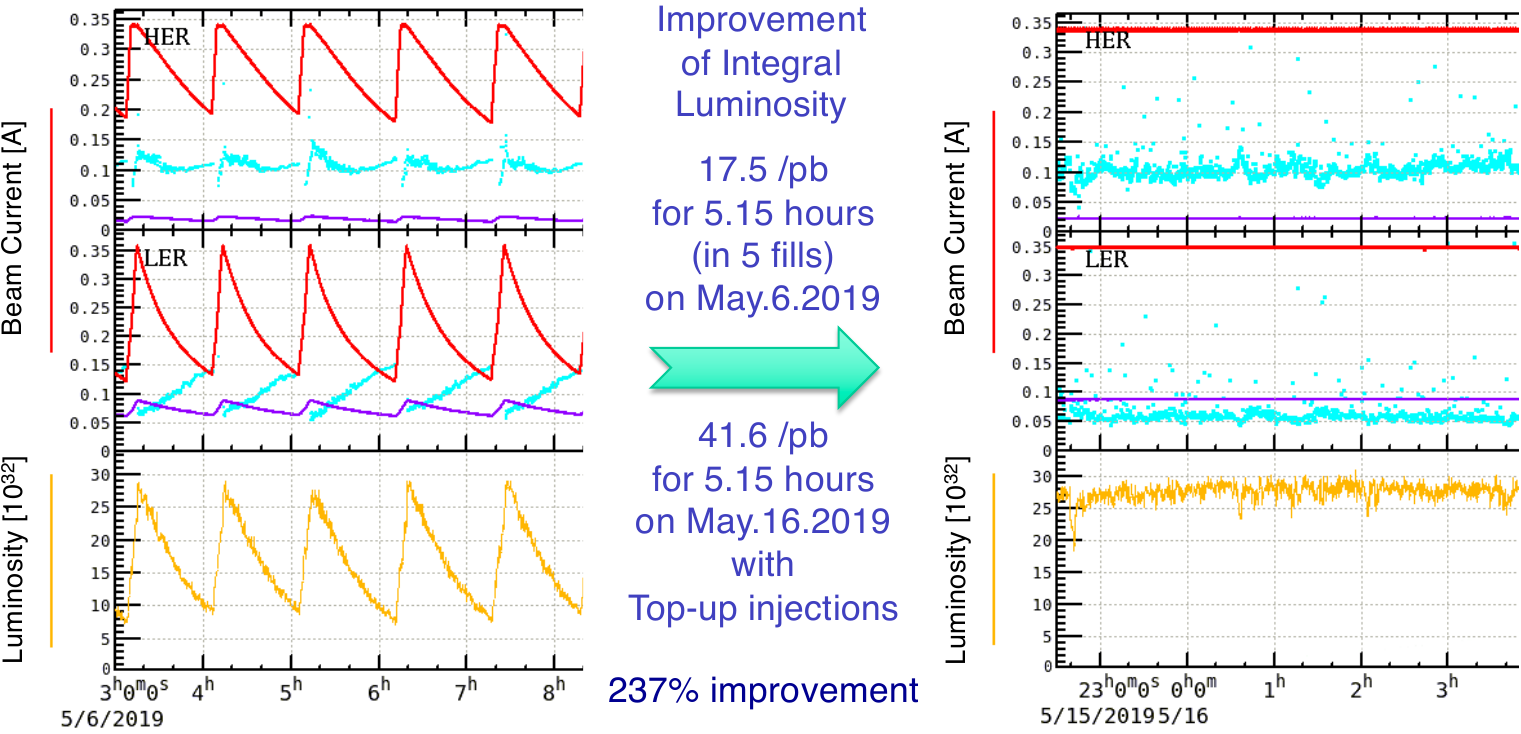}
\caption{
%%% caption Fig.5
Comparison of integrated luminosity before and after the introduction of simultaneous top-up injection.  Red lines indicate stored beam currents at storage rings, and the yellow line is the resultant luminosity that improved by 237\%.}
\label{luminosity-gain}
\end{figure}

One of the example is shown in Fig.~\ref{luminosity-gain}.  Because of the severe operational condition in SuperKEKB the simultaneous injections improved the collision performance (luminosity) by 237\%.

\section{Conclusion}

KEK injector LINAC continues multidisciplinary beam delivery with simultaneous top-up injections to support both photon science and particle physics experiments. It often carries administrative and operational negotiations to a successful conclusion to enable short-term and long-term optimizations and to enhance performances for both disciplines.

%% end Text

%%%

%
% only for "biblatex"
%
\ifboolexpr{bool{jacowbiblatex}}%
	{\printbibliography}%
	{%
	% "biblatex" is not used, go the "manual" way
	
\newcommand\etal{\emph{et~al.}}
% cancel hyperref in reference section
%% \renewcommand{\href}[2]{#2}

	} % end \ifboolexpr

\end{document}